\documentclass[twocolumn]{aastex7}

\begin{document}

\title{Are there Spectral Features in the MIRI/LRS Transmission Spectrum of K2-18b?}

\author[orcid=0000-0003-4844-9838,gname='Jake',sname='Taylor']{Jake Taylor}
\affiliation{Department of Physics, University of Oxford, Parks Rd, Oxford, OX1 3PU, UK}
\email[show]{jake.taylor@physics.ox.ac.uk}  

\begin{abstract}
Determining the composition of an exoplanet atmosphere relies on the presence of detectable spectral features. The strongest spectral features, including DMS, look approximately Gaussian. Here, I perform a suite of Gaussian feature analyses to find any statistically significant spectral features in the recently published MIRI/LRS spectrum of K2-18b \citep{Madhu2025}. In \citet{Madhu2025}, they claim a 3.4-$\sigma$ detection of spectral features compared to a flat line. In 5 out of 6 tests, I find the data preferred a flat line over a Gaussian model, with a $\chi^{2}_{\nu}$ of 1.06. When centering the Gaussian where the absorptions for DMS and DMDS peak, I find ln(B) = 1.21 in favour of the Gaussian model, with a $\chi^{2}_{\nu}$ of 0.99. With only $\sim$2-$\sigma$ in favour of Gaussian features, I conclude no strong statistical evidence for spectral features.

\end{abstract}


\section{Introduction}

The search for signs of life is a key goal of the exoplanet community. The James Webb Space Telescope (JWST) has improved our ability to study the atmospheres of temperate worlds, and many of the first attempts to study atmospheres with the telescope being of those around rocky planets. However, spectra of rocky planets are difficult to analyse, given their small features in transmission spectra, with at present no confirmed atmospheric detection \citep[e.g.][]{May2023}. On the other hand, H$_2$-rich sub-Neptunes, with their low mean molecular weight atmospheres, are readily available to study, with recent theoretical and observational work determining these could be our current best opportunity to find life elsewhere in the universe \citep{Seager2021,Seager2025}. One hypothesised class of sub-Neptune is ``Hycean'' worlds \citep{Madhu2021}, defined by a H$_2$ rich atmosphere above a global liquid water ocean. 

The planet K2-18b ($T_{\rm{eq}}$ = 254.9 $\pm$ 3.9 K, R$_p$ = 2.610 $\pm$ 0.087 R$_\oplus$, M$_p$ = 8.63 $\pm$ 1.35 M$_\oplus$ \citep{Cloutier2019,Benneke2019}) has been extensively studied, with competing hypotheses on whether it is a Hycean world \citep{Madhu2023}, mini-Neptune \citep{Wogan2024}, or a gas dwarf with a surface magma ocean \citep{Shorttle2024}. K2-18b has now been observed in transmission using 3 instrument modes of JWST: NIRISS/SOSS, NIRSpec/G395H, and MIRI/LRS. The initial near-infrared observations suggested the presence of dimethyl sulphide (DMS) \citep{Madhu2023}, but this molecule was not confirmed in a re-analysis of the same data \citep{Schmidt2025}. Recently, mid-infrared observations of K2-18b suggested the presence of DMS and/or dimethyl disulphide (DMDS) at 3-$\sigma$ significance \citep{Madhu2025}. The potential presence of DMS or DMDS has been raised as evidence for a biosphere on K2-18b \citep{Madhu2023,Madhu2025}, given that these molecules are produced by phytoplankton in Earth's oceans \citep{Alcolombri2015}. 

\citet{Madhu2025} claims to reject a flat line at 3.4-$\sigma$ significance. They compare a 16-parameter ``canonical'' model to a 1-parameter flat line model. However, this is not a nested analysis, as their flat line model is not a subset of their ``canonical'' model. It is possible to compare the Bayesian evidence of these models to determine whether one is favoured over the other, however, careful consideration of the prior ranges need to be taken, as the prior space can greatly impact the Bayesian evidence \citep{Trotta2008}. The authors do observe this behaviour; they comment that when increasing the prior range for their flat line model, they achieve a stronger significance. When using their 32-parameter ``maximal'' model\footnote{These models are being used to fit 29 data points. One should generally avoid fitting data with more free parameters than data points to avoid overfitting the data.}, they find they can only reject a flat line at 3-$\sigma$, the significance has gone down because the prior space between the ``maximal'' and ``canonical'' models are different. 

The 3-$\sigma$ detection of DMS/DMDS claimed by \citet{Madhu2025} was established using a curated ``canonical'' model as opposed to their ``maximal'' model. However, the ``canonical'' model neglects opacity contributions from other potential molecules, which consequently boosts the detection significance by neglecting alternative species in the Bayesian evidence. They performed the same tests with their "maximal" model and find they can only reject a flat line model compared to their atmosphere models without DMS/DMDS at less than the 2-$\sigma$ level. Furthermore, the authors postulate that the abundance of DMS/DMDS they obtain are consistent with $\sim$20$\times$ Earth's biogenic flux, as predicted by \citet{Tsai2024}. However, in \citet{Tsai2024} they show that ethane (C$_2$H$_6$) would be more readily detectable with this level of biogenic flux compared to DMS, yet no ethane is seen in the MIRI/LRS observations. These concerns motivate a careful reassessment of the reliability of the claimed detection of DMS and/or DMDS in K2-18b's atmosphere.

In this research note, I explore whether a flat line can be rejected by the K2-18b MIRI transmission spectrum, and therefore determine if an atmospheric signal can be claimed.

\section{Methods}

When using the Bayesian evidence to compute whether one model is favoured by a set of data compared to another model (e.g. an atmospheric model compared to a flat line), one model should ideally be a subset (`nested') of the other model with the same priors on the common parameters. In this section, I perform a nested approach to determine if a Gaussian model is statistically preferred relative to a flat line.

Following previous work \citep[e.g.][]{JWST2023,May2023}, I search for spectral features, while remaining agnostic to the particular gas causing the feature, with a Gaussian function:
\begin{equation}
    \delta_{\lambda}(A,\mu,\sigma_m,c) = A\,\text{exp}\Bigg(-\frac{(\lambda-\mu)^2)}{2\sigma^2_m}\Bigg) + c
\end{equation}
where $\delta_{\lambda}$ is the model transit depth, and $\mu$, $\sigma_m$, and $A$ are the centre, standard deviation, and amplitude of the Gaussian, respectively, and $c$ is a vertical offset. Given that the spectral shapes of DMS and DMDS are Gaussian-like (see Figure 3 of \citet{Madhu2025}), this is a reasonable approximation. I fit for the flat line using a model of the form:
\begin{equation}
    \delta_{\lambda}(c) = c\text{.}
\end{equation}

For the analysis, I only consider the observations obtained with the \texttt{JexoRes} pipeline, which is the same data shown in Figure 2 of \citet{Madhu2025}. I did perform the same analysis using the observations using the data obtained from \texttt{JexoPipe}, however, the conclusions remain similar. \footnote{The data was obtained here \url{https://osf.io/gmhw3/}}

I perform 3 different Gaussian feature tests: 1) the standard form (4 free parameters), 2) a negative Gaussian (4 free parameters), 3) a combination of the standard form and the negative Gaussian (7 free parameters). For each of these tests, I perform an additional analysis where I make informed choices of where the spectral features should be (motivated by Figure 6 of \citet{Madhu2025}, which shows that features at $\sim$ 7\,$\micron$ and $\sim$ 8.8\,$\micron$ drive the claimed detections), which therefore fixes the $\mu$ parameter. I plot the results in Figure~\ref{fig:Figure1} with the corresponding $\chi^2$, degrees of freedom, reduced $\chi^2$, and the $\ln$ Bayesian evidence obtained using \texttt{MultiNest} \citep{Feroz2009,Buchner2014}.

\section{Results}

\begin{figure*}[ht]
    \centering
    \includegraphics[width=0.99\textwidth]{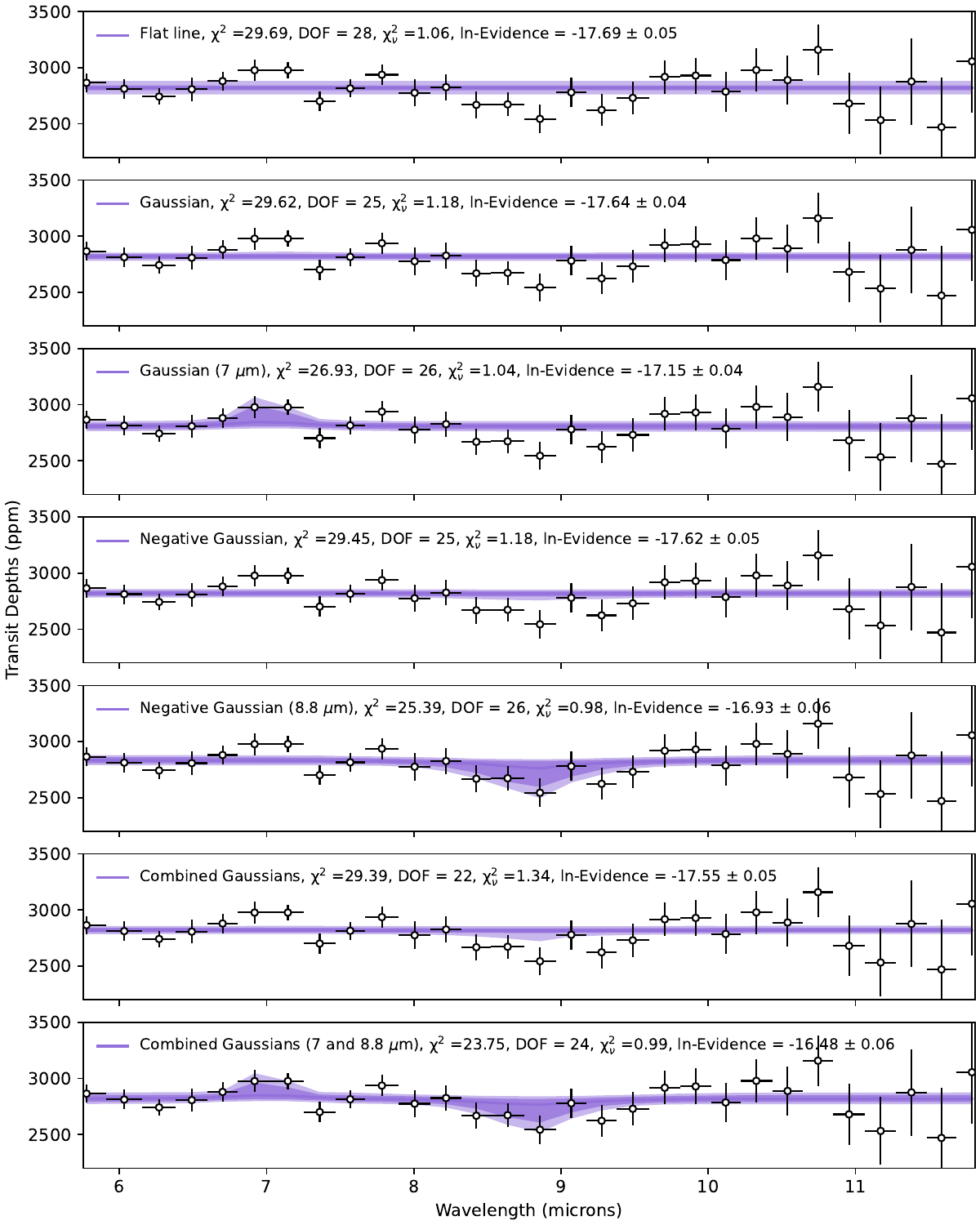}
    \caption{Model fits to the MIRI/LRS observations of K2-18b from \citep{Madhu2025}. The median, 1-$\sigma$ and 2-$\sigma$ credible region envelopes for each model fit are overlaid (purple curve and shading). The $\chi^2$, degrees of freedom, reduced $\chi^2$, and the $\ln$ Bayesian evidence values are annotated.}
    \label{fig:Figure1}
\end{figure*}

I find that all the tested models have similar ln-Bayesian evidences, indicating that a Gaussian model is not statistically preferred over a flat line. Since the difference in the ln-Evidence of a Gaussian model and the flat line model is the ln Bayes Factor ($\ln$(B)), one can assess which model can best describe the observations using the Jeffrey's scale \footnote{see Table 1 and 2 in \citet{Trotta2008}}, which defines `no evidence' ($\ln$(B) $<$ 1), `weak evidence' (1 $\leq$ $\ln$(B) $<$ 2.5), `moderate evidence' (2.5 $\leq$ $\ln$(B) $<$ 5), and `strong evidence' ($\ln$(B) $\geq$ 5). Ultimately, 5 out of 6 of the Gaussian model tests fall into the `no evidence' category according to the Jeffrey's scale, with the models with the best ln-Evidences having a Gaussian fixed to specific wavelengths. The only model marginally preferred over a flat line is the combination of a positive and negative Gaussian fixed at 7\,$\micron$ and 8.8\,$\micron$, respectively, with $\ln$(B)  = 1.21. However, even in this case, the Gaussians are only in the `weak evidence' category. When using `equivalent sigma' values \citep[e.g.][]{Trotta2008}, as commonly done in the exoplanet retrieval literature (and in \citet{Madhu2025}), this would be a `2-$\sigma$ detection' compared to the flat line model, even though there is little statistical evidence favouring the two fixed Gaussians over a flat line.

Further, when inspecting the fitted models in Figure~\ref{fig:Figure1}, one can see that the models that agnostically search for Gaussian features result in flat spectra. Only when fixing the Gaussians to specific wavelengths do I see any deviation from a flat line. This test suggests that only considering a limited set of absorbers at fixed wavelengths, as done in \citet{Madhu2025}'s ``canonical'' model, inflates the statistical evidence for an atmospheric detection.

\section{Conclusions}

I perform an independent analysis of the MIRI/LRS observations of K2-18b recently published by \citet{Madhu2025}. The specific question of this research note was to determine if there were any detectable features within the data itself, or whether the data is consistent with a flat line. I performed a Gaussian feature analysis with varying levels of complexity, as opposed to a complete atmospheric retrieval. Despite the claim of a 3.4-$\sigma$ deviation from a flat line in \citet{Madhu2025}, I find a flat line is an acceptable fit. 
Of the tested models, only one, where I use two Gaussians centred on fixed wavelengths, results in weak evidence over a flat line (ln(B) = 1.21, Bayes factor = 3.35). Therefore, there is no strong evidence for detected spectral features in K2-18b's MIRI transmission spectrum.
\label{sec:conclusions}

\begin{acknowledgments}
Thanks to Suzanne Aigrain, Louis-Philippe Coulombe, Mark Hammond, Ryan MacDonald, Peter McGill, Ben Sutlieff, and Luis Welbanks for their feedback, which improved the clarity of this manuscript.
J.T. was supported by the Glasstone Benefaction, University of Oxford (Violette and Samuel Glasstone Research Fellowships in Science 2024).
\end{acknowledgments}

\bibliography{sample7}{}
\bibliographystyle{aasjournalv7}



\end{document}